\documentclass[fleqn,usenatbib]{mnras}

\usepackage[T1]{fontenc}

\DeclareRobustCommand{\VAN}[3]{#2}
\let\VANthebibliography\thebibliography
\def\thebibliography{\DeclareRobustCommand{\VAN}[3]{##3}\VANthebibliography}

\usepackage{graphicx}	
\usepackage{amsmath}	
\usepackage{amssymb}	

\usepackage{multirow}
\usepackage[english]{babel}
\usepackage[utf8]{inputenc}
\usepackage[colorinlistoftodos]{todonotes}
\usepackage{xcolor}
\usepackage{ulem}
\usepackage{censor}

\newif\ifadd
\addfalse
\newif\ifdel
\delfalse
\newlength\nextcharwidth
\makeatletter
\renewcommand\@cenword[1]{%
  \setlength{\nextcharwidth}{\widthof{#1}}%
  \censorrule{\nextcharwidth}%
  \kern -\nextcharwidth%
  #1}
\makeatother
\newcommand{\add}[1]{\ifadd\textcolor{magenta}{#1}\else{#1}\fi}
\newcommand{\del}[1]{\ifdel\textcolor{red}{\sout{#1}}\else\fi}

\usepackage{newtxtext,newtxmath}

\newcommand{\flux}{erg s$^{-1}$ cm$^{-2}$}
\newcommand{\sdss}{Sloan Digital Sky Survey}
\newcommand{\EBV}{$E(B-V)$}
\newcommand{\tauint}{$\tau_{\text{int}}$}
\newcommand{\mcmc}{Markov Chain Monte Carlo}
\newcommand{\OIIIint}{the intrinsic [O III] $\lambda$5007}
\newcommand{\OIII}{[O III] $\lambda$5007}
\newcommand{\NIIint}{the intrinsic [N II] $\lambda$6584}
\newcommand{\NII}{[N II] $\lambda$6584}
\newcommand{\Haint}{the intrinsic H$\alpha$}

\newcommand{\metal}{$12 + \text{log\,(O/H)}$}
\newcommand{\Ha}{H${\alpha}$}
\newcommand{\Hb}{H${\beta}$}
\newcommand{\Hg}{H${\gamma}$}

\title[Re-examining Galaxies from HETDEX Pilot Survey]{Re-examining the Bayesian colour excess estimation for the local star-forming galaxies observed in the HETDEX Pilot Survey}

\author[J.-H. Shinn]{
Jong-Ho Shinn,$^{1}$\thanks{E-mail: jhshinn@kasi.re.kr}
\\
$^{1}$Korea Astronomy and Space Science Institute, 776 Daeduk-daero, Yuseong-gu, Daejeon, 34055, the Republic of Korea
}

\date{Accepted XXX. Received YYY; in original form ZZZ}

\pubyear{2022}

\begin{document}
\label{firstpage}
\pagerange{\pageref{firstpage}--\pageref{lastpage}}
\maketitle

\begin{abstract}
In my previous reanalysis of the local star-forming galaxies observed in the Hobby–Eberly Telescope Dark Energy Experiment (HETDEX) pilot survey, I reported that the overestimation of \EBV{}\add{, hence the star formation rate (SFR),} undermined the claim of new galaxy population discovery in the original study.
Herein, I re-examine whether the \EBV{} overestimation problem can be alleviated in the Bayesian parameter estimation framework by adopting scientifically motivated new priors.
I modelled the emission-line fluxes of galaxies using the strong-line method and four model parameters---metallicity \metal, nebula emission-line colour excess \EBV{}, intrinsic \OIII{} line flux and intrinsic \NII{} line flux.
Based on mock-data tests, I found that all input values can be recovered within and around the 1-$\sigma$ credible interval by adopting suitable priors for \OIIIint{} and \NII{} line fluxes: the inverse gamma distribution reflecting the logical constraint that an intrinsic emission-line flux must exceed the observed (reddened) emission-line flux.
The mock-data tests were performed for two metallicity calibrations, three colour excess input values [\EBV{} = 0.1, 0.3 and 0.5] and two metallicity input values [\metal{} = 8.0 and 8.5].
\add{I also found that the new prior can diminish the SFR overestimation eightfold.}
This study demonstrates how the Bayesian parameter estimation can achieve more accurate estimates with no further observations when the likelihood does not constrain the model parameters correctly.
\end{abstract}




\section{Introduction} \label{intro}
When an incident occurs in our daily lives, we often try to trace its cause based on prior knowledge.
The same reasoning is instrumental in scientific research because limited data and background information are usually available to scientists.
The Bayesian probability theory provides a coherent conceptual framework for this type of reasoning \citep{Sivia_2006_book,Toussaint_2011_RvMP_83_943,Sharma_2017_ARA&A_55_213}.
In this theory, \textit{probability} is defined as the degree of belief in contrast to the classical definition of limiting relative frequencies \citep{Wasserman_2004_book}.
Furthermore, Bayesian statistical inference is based on Bayes' theorem, which relates four conditional probabilities as follows \citep{Sivia_2006_book,Toussaint_2011_RvMP_83_943,Sharma_2017_ARA&A_55_213},
\begin{eqnarray} \label{eq-bayes}
    p(H|D,I) & = &\frac{p(D|H,I)\,p(H|I)}{p(D|I)},
\end{eqnarray}
or
\begin{eqnarray} \label{eq-name}
    \text{Posterior\footnotemark[1]} & = & \frac{\text{Likelihood\footnotemark}\times\text{Prior\footnotemark[1]}}{\text{Evidence\footnotemark[1]}},
\end{eqnarray}
\footnotetext{`Posterior' and `Prior' are probability density functions; hence, they must be unity when summed over the parameters ($H$). `Likelihood' is the joint probability density of the data ($D$), not a probability density function. `Evidence' is a normalising constant equal to the integration of `Likelihood'$\times$`Prior' over all parameters ($H$). In this sense, `Evidence' is also called `marginal likelihood'. See \cite{Wasserman_2004_book} and \cite{Sivia_2006_book} for more details.}
where $p(\cdot|\cdot)$, $H$, $D$ and $I$ denote the conditional probability, hypothesis, data and background information, respectively.
Bayesian inference covers diverse topics such as parameter estimation, model comparison and hierarchical modelling \citep{Toussaint_2011_RvMP_83_943,Sharma_2017_ARA&A_55_213} and has established a presence in astronomy over the past two decades \citep{Sharma_2017_ARA&A_55_213}.
A few application examples are as follows: parameter estimation \citep{Gregory_2005_ApJ_631_1198,Miller_2007_MNRAS_382_315,Steiner_2010_ApJ_722_33,Brown_2011_AJ_142_112,Buchner_2014_A&A_564_A125}, model comparison \citep{Farr_2011_ApJ_741_103,Diaz_2014_MNRAS_441_983,Buchner_2014_A&A_564_A125,PlanckCollaboration_2018_arXiv_07_6211} and hierarchical modelling \citep{Wolfgang_2015_ApJ_806_183,Martinez_2015_MNRAS_451_2524,Birrer_2019_MNRAS_484_4726}.

Although Bayesian inference has garnered increasing attention in astronomy, studies have also reported the inappropriate use of Bayesian analysis \citep[e.g.][]{Starck_2013_A&A_552_A133,DAntona_2018_NatAs__,Tak_2018_MNRAS_481_277,Cameron_2020_NatAs_4_132}.
In particular, \cite{Tak_2018_MNRAS_481_277} emphasised that the posterior must be unity when summed over the parameters, noting the risk of naive use of flat priors in the entire parameter space to reflect a lack of prior knowledge.
This requirement on the posterior is called \textit{posterior propriety}.
If a prior diverges when integrated over its entire parameter space, i.e. $\int p(\theta)d\theta=\infty$ for parameter $\theta$, it is referred to as an \textit{improper prior} \citep{Wasserman_2004_book}.
For example, a flat (uniform) prior over (0, $\infty$) or ($-\infty$, $\infty$) is an improper prior.
\cite{Tak_2018_MNRAS_481_277} warned that this improper prior does not guarantee the posterior propriety because the posterior propriety requires $\int p(D|\boldsymbol\Theta)\,p(\boldsymbol\Theta)\, d\boldsymbol\Theta$, where $\boldsymbol\Theta$ denotes all parameters, to be finite based on Bayes' theorem [eq.~(\ref{eq-bayes})].
To establish this posterior propriety, one of the following conditions must be met \citep{Hobert_1996_J.Am.Stat.Assoc._91_1461}: (1) the posterior propriety must be analytically proven when improper priors are used or (2) \textit{proper priors} must be jointly used for $\boldsymbol\Theta$.
Analytically proving the posterior propriety is difficult in most astronomy scenarios because of large parameter dimensions or model complexity \citep{Tak_2018_MNRAS_481_277}.
Therefore, \cite{Tak_2018_MNRAS_481_277} suggested some scientifically motivated proper priors such as generalised Gaussian, inverse gamma and multiply-broken power-law priors.

Recently, \cite{Indahl_2019_ApJ_883_114} studied local ($z<0.15$) star-forming galaxies in terms of gas metallicity, stellar mass and star formation rate (SFR), using the Bayesian analysis; I reanalysed their data by carefully considering the analysis method \citep{Shinn_2020_MNRAS_499_1073}.
\cite{Indahl_2019_ApJ_883_114} and I estimated the gas metallicity from the emission line ratios using the strong-line method, which yields an empirical relation between metallicity and ratios of strong emission lines \citep{Maiolino_2019_A&ARv_27_3}.
We independently modelled emission lines based on these empirical relations.
We performed the Bayesian parameter estimation on three or four model parameters: metallicity expressed in terms of the oxygen abundance \metal, nebular emission-line colour excess \EBV{}, intrinsic \OIII{} line flux and intrinsic \Ha{} or intrinsic \NII{} line flux (introduced when needed).
I achieved the posterior propriety by adopting proper priors.

However, some points must be reconsidered in my priors.
First, the upper limit of the emission-line flux ($10^{-8}$ \flux{}) is arbitrary and does not have any scientific basis; I employed a sufficiently large value.
Second, the intrinsic emission-line flux is unlikely to achieve a uniform probability over the range because fewer galaxies would show higher intrinsic emission-line fluxes.
Third, a new scientifically motivated prior other than a flat prior may yield improved parameter estimation results.
I found that \cite{Indahl_2019_ApJ_883_114} probably overestimated \EBV{} for galaxies with low-\EBV{} in their Bayesian analysis setup, which can afford a factor of five overestimations of SFR \citep{Shinn_2020_MNRAS_499_1073}.
If a new prior for the intrinsic emission-line flux can reduce this overestimation of \EBV{}, a more accurate SFR estimation can be achieved based on the reddening-corrected luminosity of emission lines.
This point is worthy of further investigation, considering that my solutions suggested in \cite{Shinn_2020_MNRAS_499_1073} for the \EBV{} overestimation problem require reducing the target galaxy number or additional observation times.

Herein, I re-examine the \EBV{} overestimation problem based on the Bayesian analysis results \cite{Shinn_2020_MNRAS_499_1073} reported using new scientifically motivated proper priors.
I generated mock data and tested the recovery of the input values when the new priors were adopted.
As the new prior, I adopted the inverse gamma prior, one of the proper priors proposed by \cite{Tak_2018_MNRAS_481_277}, for the intrinsic \OIII{} and intrinsic \NII{} line fluxes.
I adjusted the priors to reflect the information that the reddened emission-line fluxes of the mock data are the lower limits of the intrinsic emission-line fluxes.
I found that the \EBV{} overestimation can be considerably reduced when using well-suited proper priors; all input values for the mock data fall within or around the 1-$\sigma$ credible limits of the corresponding parameter posteriors.

\section{Previous Modelling for Emission-line Flux of Star-forming Galaxies and \EBV{} Overestimation} \label{model-EBV}
The study performed by \cite{Indahl_2019_ApJ_883_114} on local star-forming galaxies and my reanalysis of their data are based on emission-line flux modelling for the gas metallicity estimation.
In this section, I describe the emission-line flux model used in the literature \citep{Indahl_2019_ApJ_883_114,Shinn_2020_MNRAS_499_1073} and the relevant backgrounds.

\cite{Indahl_2019_ApJ_883_114} modelled emission-line fluxes to estimate the gas metallicity of the local ($z<0.15$) star-forming galaxies observed in the pilot survey for the Hobby–Eberly Telescope Dark-Energy Experiment (HETDEX, \citealt{Hill_2008_inproca}) called the HETDEX pilot survey (HPS, \citealt{Adams_2011_ApJS_192_5}).
The HETDEX is a blind spectroscopic survey based on an integral-field-unit spectrograph.
It surveys an area of 420/4.5 deg$^2$ with coverages of $\sim3500-5500$ \AA{} and a resolution of $\sim5.7$ \AA{}.
For the HPS, a similar wavelength coverage ($\sim3500-5800$ \AA{}) and spectral resolution ($\sim5$ \AA{}) are considered; however, the survey area is smaller ($\sim163$ arcmin$^2$).
\cite{Indahl_2019_ApJ_883_114} secured 29 galaxies with at least three emission lines---[O II] $\lambda3727$, [O III] $\lambda5007$ and H$\beta$---from the HPS dataset and follow-up observations\add{{ }with the second-generation Low Resolution Spectrograph (LRS2; \citealt{Chonis_2016_inbook}) of the Hobby–Eberly Telescope \citep[][]{Ramsey_1998_inproc,Hill_2021_AJ_162_298}}.

To estimate the gas metallicity from the observed emission lines, \cite{Indahl_2019_ApJ_883_114} employed the strong-line method, in which empirical relations between the metallicity \metal{} and the flux ratio of strong emission lines (i.e. the lines easier to detect) are used.
The strong-line method was invented for a more straightforward metallicity estimation because more accurate methods, such as electron temperature and recombination line methods, exploit the weaker metallic emission lines, which are weaker than the Balmer lines by a factor of $\sim10-10^4$ \citep{Maiolino_2019_A&ARv_27_3}.
Among numerous calibrations of strong lines based on the electron temperature method, a photoionisation model, or both \citep[for details, refer to][]{Maiolino_2019_A&ARv_27_3}, \cite{Indahl_2019_ApJ_883_114} adopted the calibration employed by \cite{Maiolino_2008_A&A_488_463}.
\cite{Maiolino_2008_A&A_488_463} calibrated the gas metallicity with local star-forming galaxies ($z\sim0$) by separately using the electron temperature method and photoionisation model in two metallicity ranges.
\cite{Indahl_2019_ApJ_883_114} used the following three ratios 
\begin{equation} \label{R23}
    \text{R23}=\frac{\text{[O II]}\,\lambda3727 + \text{[O III]}\,\lambda4959 + \text{[O III]}\,\lambda5007}{\text{H}\beta},
\end{equation}
\begin{equation} \label{O32}
    \text{O32}=\frac{\text{[O III]}\,\lambda5007}{\text{[O II]}\,\lambda3727},
\end{equation}
\begin{equation} \label{N2}
    \text{N2}=\frac{\text{[N II]}\,\lambda6584}{\text{H}\alpha},
\end{equation}
and followed the approach reported by \cite{GrasshornGebhardt_2016_ApJ_817_10} when modelling emission-line fluxes.
\cite{Indahl_2019_ApJ_883_114} modelled the intrinsic emission-line fluxes using the three line-ratios and then calculated the reddened emission-line flux using the Calzetti attenuation curve \citep{Calzetti_2000_ApJ_533_682}.
The model parameters are metallicity \metal{}, nebular emission-line colour excess \EBV{} and intrinsic \OIII{} line flux; \cite{Indahl_2019_ApJ_883_114} added another parameter, namely, \Haint{} line flux when they additionally used the [N II] $\lambda6584$ line for the metallicity estimation.
\cite{Indahl_2019_ApJ_883_114} fixed the ratio of [O III] $\lambda$5007 to [O III] $\lambda$4959 at 2.98 by referring to \cite{Storey_2000_MNRAS_312_813}.

\cite{Indahl_2019_ApJ_883_114} estimated the model parameters by sampling the posteriors with the \mcmc{} (MCMC) method \citep[for more details on the MCMC method, refer to][]{Sharma_2017_ARA&A_55_213,Hogg_2018_ApJS_236_11}.
They employed the \textsf{emcee} package \citep{Foreman-Mackey_2013_PASP_125_306,Foreman-Mackey_2019_JOSS_4_1864} for the MCMC sampling.
Their log-likelihood expression is shown below:
\begin{equation}
    \text{ln}\,\mathcal{L} \sim -\frac{1}{2}\sum_{l}^{} \frac{(x_{\text{obs},l}-x_{\text{mod},l})^2}{\sigma_{\text{obs},l}^2+\sigma_{\text{mod},l}^2},
    \label{eq-L}
\end{equation}
where $l$ denotes the different emission lines over which the fraction is summed; $x$ and $\sigma$ represent the emission-line flux and its uncertainty, respectively; and the subscripts `obs' and `mod' denote the corresponding values from the observations and the model, respectively.
This likelihood includes the uncertainty in the model emission-line flux ($\sigma_{\text{mod},l}$), which is modelled to mimic the scatter of the line ratios for a given metallicity \citep[see][]{Maiolino_2008_A&A_488_463}.
\cite{Indahl_2019_ApJ_883_114} adopted flat priors for the \metal{}, intrinsic \OIII{} line flux and intrinsic \NII{} line flux while adopting a normal (Gaussian) prior for \EBV{}.

In my previous reanalysis of the data reported by \cite{Indahl_2019_ApJ_883_114} \citep{Shinn_2020_MNRAS_499_1073}, I followed the model setup employed by \cite{Indahl_2019_ApJ_883_114} with two exceptions.
First, I increased the uncertainty to adequately cover the scatter of the line ratio seen in the data points used for the metallicity calibration by \cite{Maiolino_2008_A&A_488_463}.
Second, I set \NIIint{} line flux as a model parameter instead of \Haint{} line flux when [N II] $\lambda6584$ lines were used for the metallicity estimation.
Then, I performed two steps not conducted in the analysis by \cite{Indahl_2019_ApJ_883_114}.
The first step involved the convergence monitoring of the MCMC sampling.
The MCMC method randomly samples the target distribution; hence, the sampling convergence should be monitored to assess the closeness of the sample to the target distribution.
I monitored the integrated autocorrelation time (\tauint)\footnote{
The integrated autocorrelation time (\tauint) is expressed as follows \citep{Sharma_2017_ARA&A_55_213}:
\begin{equation} \label{eq-iat}
    \tau_{\text{int}}=\sum^{\infty}_{t=-\infty}\rho_{xx}(t)
    \text{, where } \rho_{xx}(t)=\frac{\mathbb{E}[(x_i-\bar{x})(x_{i+t}-\bar{x})]}{\mathbb{E}[(x_i-\bar{x})^2]},
\end{equation}
where $\rho_{xx}$ represents the autocorrelation function for the sequence $\{x_i\}$, $t$ denotes the time difference---or distance---between two points in the sequence $\{x_i\}$, $\bar{x}$ represents the mean of the sequence $\{x_i\}$ and $\mathbb{E}\left[\cdot\right]$ denotes the expectation value.
}, which reflects the degree of correlation of the sample with itself.
Owing to this correlation, the MCMC sampling affords only N/\tauint{} independent samples, where N represents the total sample length \citep{Sharma_2017_ARA&A_55_213}; this number of independent samples is called the effective sample size (ESS).
I stopped the MCMC sampling when all parameters achieved an ESS of $>2000$.
This ESS threshold is greater than the value of 1665 required to determine a 0.025 quantile to within $\pm$0.0075 with a probability of 0.95, which corresponds to a $\sim$10\% error in the 0.025 quantiles for light-tailed (normal) or moderate-tailed (Student's $t_{\nu=4}$) distributions \citep{Raftery_1992_inproc}.
The second step involved the mock-data test for the parameter estimation.
I generated mock data from the model and assessed the extent of the recovery of the input values for the mock data via the parameter estimation process.
From this test, I found that the parameter \EBV{} was overestimated for galaxies with low-\EBV{} because of the model uncertainty, attributed to the scatters in the metallicity calibration (Fig.~\ref{fig-model_test}), possibly resulting in the SFR overestimation up to a factor of five.

To overcome this \EBV{} overestimation problem, I suggested an independent \EBV{} estimation using the Balmer decrement of the \Ha-\Hb{} or \Hb-\Hg{} pair \citep{Shinn_2020_MNRAS_499_1073}.
However, this solution demands either a reduction in the number of target galaxies or follow-up observations because the \Hg{} line is weaker than the \Ha{} or \Hb{} line, and the HETDEX survey does not cover \Ha{}.
This study explores whether the \EBV{} overestimation problem can be prevented using scientifically motivated priors.

\section{Analysis and Results}
In this study, I aim to determine whether new scientifically motivated proper priors can alleviate the \EBV{} overestimation problem previously reported by me \citep{Shinn_2020_MNRAS_499_1073} (section \ref{model-EBV}).
To begin with, I used a slightly modified likelihood to more accurately consider the effects of the model uncertainty in the likelihood on the posterior.
The modification of the likelihood is described in section \ref{ana-res-modL}.
Then, the mock-data tests performed using the new priors and a more recent metallicity calibration are presented in sections \ref{ana-res-mock} and \ref{ana-res-cal}, respectively.

\subsection{Modification of the Likelihood} \label{ana-res-modL}
\cite{Indahl_2019_ApJ_883_114} included the uncertainty of the model emission-line flux in the likelihood by adding it in quadrature [refer to $\sigma_{\text{mod},l}$ in eq.~(\ref{eq-L})], and I followed their approach \citep{Shinn_2020_MNRAS_499_1073}.
This log-likelihood expression is derived from the likelihood function, whose probability density function (PDF) is assumed to be a normal distribution.
Therefore, the likelihood has a $1/\sigma$ factor from the normal distribution; however, this factor is usually ignored because it is a constant value when $\sigma$ is obtained only from the data.
However, in our case, $\sigma$ is obtained from both the data and model and the $1/\sigma$ factor is no longer constant.
Therefore, rationally, this factor should be included in the likelihood.
I modified the likelihood to include the $1/\sigma$ factor; then, the log-likelihood expression becomes 
\begin{equation}
    \text{ln}\,\mathcal{L} \sim -\frac{1}{2}\left[\sum_{l}^{} \left\{ \frac{(x_{\text{obs},l}-x_{\text{mod},l})^2}{\sigma_{\text{obs},l}^2+\sigma_{\text{mod},l}^2} + \text{ln}\,({\sigma_{\text{obs},l}^2+\sigma_{\text{mod},l}^2}) \right\} \right].
    \label{eq-modL}
\end{equation}
I used this modified likelihood in all subsequent analyses in this study.

\subsection{Mock-Data Test using the Inverse Gamma Priors \label{ana-res-mock}}

My emission-line flux model has four parameters, two of which are \OIIIint{} and intrinsic \NII{} line fluxes (section \ref{model-EBV}).
These emission lines are emanated from the target galaxies, and their reddened emission-line flux affected by dust attenuation is measured.
Therefore, the reddened emission-line flux can be treated as the lower limit for the intrinsic emission-line flux.
Based on this, I revised the priors for \OIIIint{} and intrinsic \NII{} line fluxes to achieve lower limits.
As the parameter values increase, these priors should gradually approach zero because few galaxies are observed at a high-value domain.

\begin{figure}
    \center{
        \includegraphics[scale=0.5]{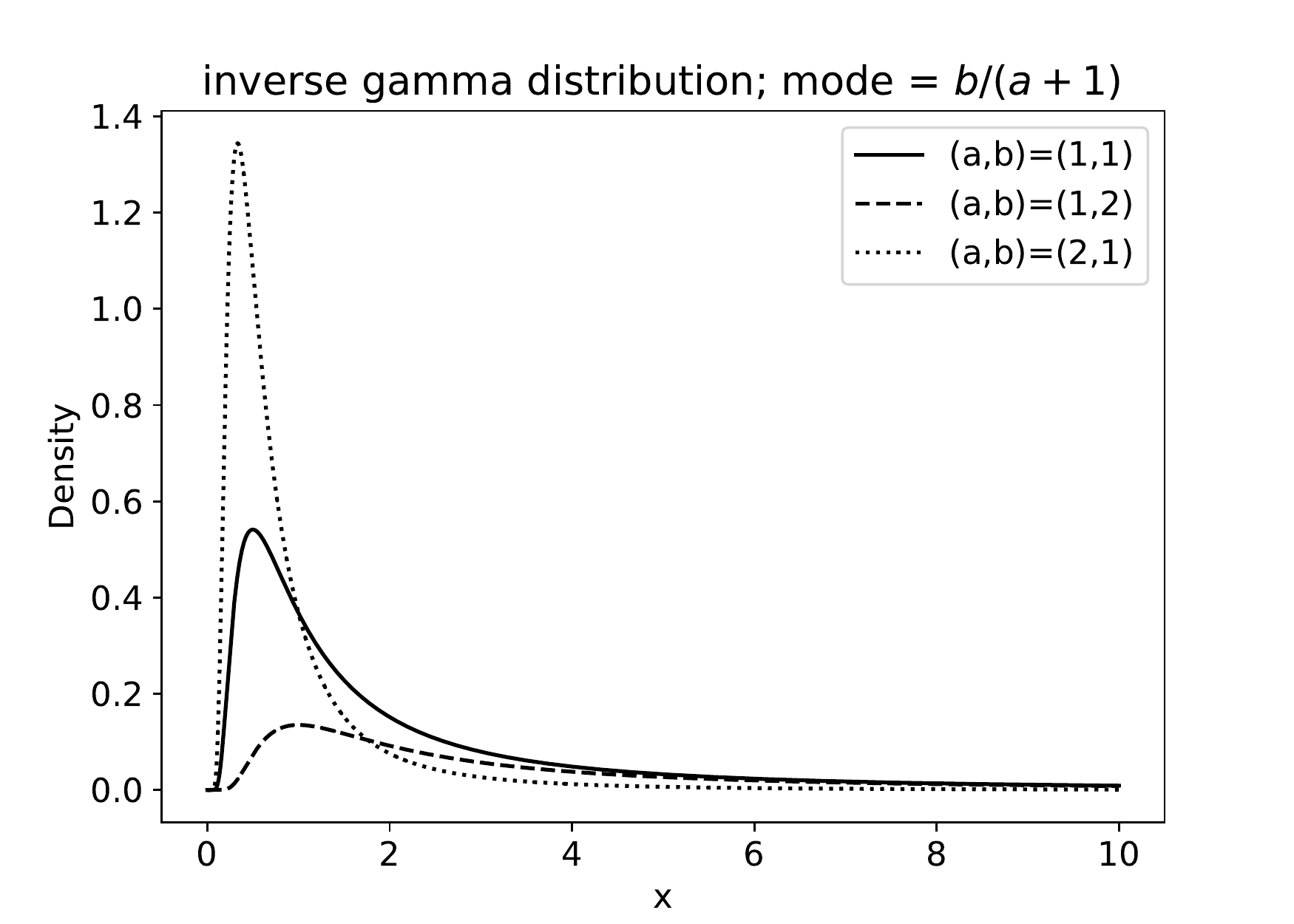}
    }
    \caption{Three examples of the inverse gamma distribution for different pairs of shape and scale parameters, i.e. $(a, b)$. The mode of the distribution is at $x=b/(a+1)$.}
    \label{fig-invgam}
\end{figure}

For this, I adopted the inverse gamma distribution. \cite{Tak_2018_MNRAS_481_277} recommended this distribution as a proper prior that can be used when the parameter achieves only positive real values with a soft lower bound.
The inverse gamma distribution---$f(x;a,b)\sim x^{-a-1}\textrm{exp}(-b/x)$---has shape ($a>0$) and scale ($b>0$) parameters, and its mode is at $x=b/(a+1)$.
Fig.~\ref{fig-invgam} shows some examples of the inverse gamma distribution for different ($a,b$) pairs.
The examples achieve a single peak and approach zero as $x$ increases.
Note that the values at the range smaller than the mode is low; in other words, the mode can be treated as a soft lower bound.

I executed mock-data tests to assess the extent of the recovery of the input values for the mock data using the inverse gamma priors.
I adopted two inverse gamma priors of $a=1$ and 2 to assess the effects of different shape parameters on the recovery of the input values.
The scale parameter $b$ was adjusted such that the mode of the prior is equal to the reddened flux of the \OIII{} and \NII{} lines.
For comparison, I also performed the same mock-data test by adopting flat priors for \OIIIint{} and intrinsic \NII{} line fluxes.
Table \ref{tbl-priors} lists the specifications of the adopted priors for all four model parameters.
For the mock-data tests shown in this subsection, I adopted the metallicity calibration reported by \cite{Maiolino_2008_A&A_488_463}, the same one adopted by \cite{Indahl_2019_ApJ_883_114} and \cite{Shinn_2020_MNRAS_499_1073}.

\begin{table}
    \centering
    \caption{Adopted priors for the mock-data tests \label{tbl-priors}}
    \begin{tabular}{ccc}
        \hline
        \multirow{2}{*}{Parameter} & \multicolumn{2}{c}{Metallicity calibration} \\
        \cline{2-3}
        & {\cite{Maiolino_2008_A&A_488_463}} & {\cite{Curti_2020_MNRAS_491_944}} \\
        \hline
        \metal{}    &   Flat; (7.1, 9.1)$^a$    & Flat; (7.6, 8.9)$^a$  \\
        \EBV{} &   Flat; (0.00, 0.79)$^b$    & same  \\
        Intrinsic \OIII{}  &   Flat; (0., $10^{-8}$)$^c$   & \multirow{2}{*}{same}  \\
        line flux           & Inverse gamma(a, b)$^d$ & \\
        Intrinsic \NII{}  &   Flat; (0., $10^{-8}$)$^c$   & \multirow{2}{*}{same} \\
        line flux           & Inverse gamma(a, b)$^d$ & \\
        \hline
        \multicolumn{3}{l}{\parbox{8cm}{$^a$ This corresponds to the valid range of the corresponding metallicity calibration.}} \\
        \multicolumn{3}{l}{\parbox{8cm}{$^b$ The maximum value corresponds to $\mu+3\sigma$ of normal distribution adopted by \cite{Indahl_2019_ApJ_883_114}, where $(\mu,\sigma)=(0.295,0.165)$.}} \\
        \multicolumn{3}{l}{$^c$ Unit is \flux.} \\
        \multicolumn{3}{l}{\parbox{8cm}{$^d$ The shape parameter `a' is 1 or 2, and the scale parameter `b' is adjusted such that its mode corresponds to the reddened emission-line flux. Refer to Fig.~\ref{fig-invgam} for the form of the inverse gamma distribution.}} \\
    \end{tabular}
\end{table}

The input values for the mock data are listed in Table \ref{tbl-results}.
The \metal{} value was set to 8.0 because it is in the middle of the metallicity calibration coverage (Table \ref{tbl-priors}).
The intrinsic \OIII{} and intrinsic \NII{} line fluxes were arbitrarily set at a magnitude similar to the typically observed emission-line fluxes.
Three values were selected for \EBV{}, the main interest of this study, to assess the extent to which the input value is recovered in the probable range of \EBV{}: \EBV{} = 0.1, 0.3 and 0.5.
The signal-to-noise ratio of the mock data was set to 10, the typical value obtained from the HPS data \citep[][]{Shinn_2020_MNRAS_499_1073}.

\begin{table*}
    \centering
    \caption{Input values for the mock-data test and estimates from the posterior distributions (median and 1-$\sigma$ credible interval) \label{tbl-results}}
    \begin{tabular}{cccccccc}
        \hline
        \multirow{2}{*}{Parameter} & \multirow{2}{*}{Input value} & \multicolumn{2}{c}{prior: flat$^a$} &  \multicolumn{2}{c}{prior: invGau a1$^a$} & \multicolumn{2}{c}{prior: invGau a2$^a$} \\
        \cline{3-8}
        &   & {metcal: M$^b$} & {metcal: C$^b$} & {metcal: M$^b$} & {metcal: C$^b$} & {metcal: M$^b$} & {metcal: C$^b$} \\
        \hline
        \multicolumn{8}{c}{Input \EBV{} = 0.1}\\
        \hline
        12 + log(O/H) & 8.00  & ${8.07}^{+0.08}_{-0.09}$ & ${8.09}^{+0.06}_{-0.08}$ & ${8.00}^{+0.09}_{-0.10}$ & ${8.02}^{+0.08}_{-0.09}$ & ${7.99}^{+0.09}_{-0.10}$ & ${8.01}^{+0.08}_{-0.09}$\\
\EBV{} & 0.10  & ${0.61}^{+0.13}_{-0.20}$ & ${0.60}^{+0.13}_{-0.21}$ & ${0.23}^{+0.17}_{-0.14}$ & ${0.22}^{+0.16}_{-0.13}$ & ${0.16}^{+0.13}_{-0.10}$ & ${0.16}^{+0.13}_{-0.10}$\\
Intrinsic [O III] line flux$^c$ & 1.00  & ${8.15}^{+5.93}_{-4.61}$ & ${7.93}^{+5.87}_{-4.65}$ & ${1.68}^{+1.80}_{-0.73}$ & ${1.62}^{+1.53}_{-0.69}$ & ${1.27}^{+0.95}_{-0.43}$ & ${1.27}^{+0.94}_{-0.43}$\\
Intrinsic [N II] line flux$^c$ & 2.00  & ${9.45}^{+4.98}_{-4.44}$ & ${9.35}^{+4.85}_{-4.56}$ & ${2.92}^{+2.09}_{-1.01}$ & ${2.84}^{+1.86}_{-0.96}$ & ${2.36}^{+1.23}_{-0.64}$ & ${2.36}^{+1.22}_{-0.63}$\\

        \hline
        \multicolumn{8}{c}{Input \EBV{} = 0.3}\\
        \hline
        12 + log(O/H) & 8.00  & ${8.04}^{+0.08}_{-0.09}$ & ${8.06}^{+0.07}_{-0.08}$ & ${7.98}^{+0.10}_{-0.11}$ & ${7.99}^{+0.09}_{-0.10}$ & ${7.96}^{+0.10}_{-0.11}$ & ${7.97}^{+0.09}_{-0.10}$\\
\EBV{} & 0.30  & ${0.66}^{+0.09}_{-0.18}$ & ${0.65}^{+0.10}_{-0.19}$ & ${0.28}^{+0.21}_{-0.16}$ & ${0.27}^{+0.18}_{-0.15}$ & ${0.17}^{+0.13}_{-0.10}$ & ${0.18}^{+0.14}_{-0.11}$\\
Intrinsic [O III] line flux$^c$ & 1.00  & ${4.47}^{+2.15}_{-2.33}$ & ${4.24}^{+2.33}_{-2.30}$ & ${0.89}^{+1.22}_{-0.43}$ & ${0.87}^{+1.02}_{-0.41}$ & ${0.57}^{+0.42}_{-0.20}$ & ${0.60}^{+0.47}_{-0.22}$\\
Intrinsic [N II] line flux$^c$ & 2.00  & ${6.07}^{+2.20}_{-2.58}$ & ${5.79}^{+2.27}_{-2.56}$ & ${1.83}^{+1.64}_{-0.72}$ & ${1.79}^{+1.40}_{-0.68}$ & ${1.31}^{+0.68}_{-0.36}$ & ${1.37}^{+0.73}_{-0.40}$\\

        \hline
        \multicolumn{8}{c}{Input \EBV{} = 0.5}\\
        \hline
        12 + log(O/H) & 8.00  & ${8.00}^{+0.09}_{-0.10}$ & ${8.02}^{+0.08}_{-0.09}$ & ${7.95}^{+0.10}_{-0.12}$ & ${7.96}^{+0.10}_{-0.10}$ & ${7.95}^{+0.11}_{-0.12}$ & ${7.94}^{+0.10}_{-0.10}$\\
\EBV{} & 0.50  & ${0.68}^{+0.08}_{-0.17}$ & ${0.68}^{+0.08}_{-0.16}$ & ${0.31}^{+0.21}_{-0.17}$ & ${0.32}^{+0.19}_{-0.17}$ & ${0.20}^{+0.15}_{-0.12}$ & ${0.21}^{+0.15}_{-0.12}$\\
Intrinsic [O III] line flux$^c$ & 1.00  & ${2.12}^{+0.84}_{-1.08}$ & ${2.14}^{+0.83}_{-1.06}$ & ${0.45}^{+0.65}_{-0.23}$ & ${0.46}^{+0.57}_{-0.23}$ & ${0.28}^{+0.24}_{-0.11}$ & ${0.29}^{+0.26}_{-0.11}$\\
Intrinsic [N II] line flux$^c$ & 2.00  & ${3.45}^{+1.04}_{-1.46}$ & ${3.49}^{+1.04}_{-1.41}$ & ${1.09}^{+1.01}_{-0.44}$ & ${1.11}^{+0.92}_{-0.45}$ & ${0.77}^{+0.45}_{-0.24}$ & ${0.80}^{+0.48}_{-0.25}$\\

        \hline
        \multicolumn{8}{l}{\parbox{13cm}{$^a$ The prior types for \OIIIint{} and intrinsic \NII{} line fluxes. The `flat', `invGam a1' and `invGam a2' indicate the flat, inverse gamma with $a=1$ and inverse gamma with $a=2$ priors, respectively.}} \\
        \multicolumn{8}{l}{\parbox{13cm}{$^b$ This indicates the adopted metallicity calibration functions. `M' denotes \cite{Maiolino_2008_A&A_488_463}, and `C' denotes \cite{Curti_2020_MNRAS_491_944}. }} \\
        \multicolumn{8}{l}{$^c$ Unit is $10^{-16}$ \flux.} \\
    \end{tabular}
\end{table*}

Then, I performed the Bayesian parameter estimation on the four model parameters by sampling the posterior [eqs.~(\ref{eq-bayes}), (\ref{eq-name}) and (\ref{eq-modL})]\footnote{`Evidence' is a certain constant, which is equal to $\int p(D|\boldsymbol\Theta)\,p(\boldsymbol\Theta)\, d\boldsymbol\Theta$ \citep[see][]{Sharma_2017_ARA&A_55_213} and independent of $\Theta$; hence, it can be ignored for parameter estimation.}, as performed previously \citep{Shinn_2020_MNRAS_499_1073}.
First, I set 32 points around the mode of the posterior using a global optimisation method called differential evolution \citep{Storn_1997_J.GlobalOptim._11_341}.
Second, starting from these 32 points (i.e. 32 walkers), I sampled the posterior using the MCMC method.
I employed the affine-invariant ensemble sampler called
\textsf{emcee} \citep[][]{Foreman-Mackey_2013_PASP_125_306,Foreman-Mackey_2019_JOSS_4_1864} and used the stretch move \citep[][]{Goodman_2010_CAMCoS_5_65} with the stretch scale parameter $a$ = 2.
During the MCMC sampling, I assessed the sampling convergence by monitoring the samples' integrated autocorrelation time (\tauint) [eq.~(\ref{eq-iat})].
I excluded the samples in the `burn-in'\footnote{`Burn-in' indicates the process where the sampling walker moves from its initial value to the area where it samples repeatedly. See \cite{Brooks_2011_book} and \cite{Hogg_2018_ApJS_236_11} for more about ‘burn-in’.} phase when monitoring the convergence and performing the parameter estimation.
I stopped the MCMC sampling when the ESSs for all parameters were > 2000, considering that the sampling achieved a sufficient convergence; the ESS is equal to N/\tauint{} (N is the total sample length) and a measure of the number of independent samples that can be obtained via the sampling \citep{Sharma_2017_ARA&A_55_213}.
Lastly, to determine whether any local maxima were missed during the sampling, I performed a check-up sampling using a fivefold higher stretch-scale parameter ($a=10$), enabling a more extensive search of parameter ranges.
I started the check-up sampling at the last walker positions of the previous sampling and performed the sampling for 1000 iterations.
In all the check-up samplings, no local maxima were achieved.
More details on the posterior sampling are described in \cite{Shinn_2020_MNRAS_499_1073}.

\begin{figure*}
    \center{
        \includegraphics[scale=0.55]{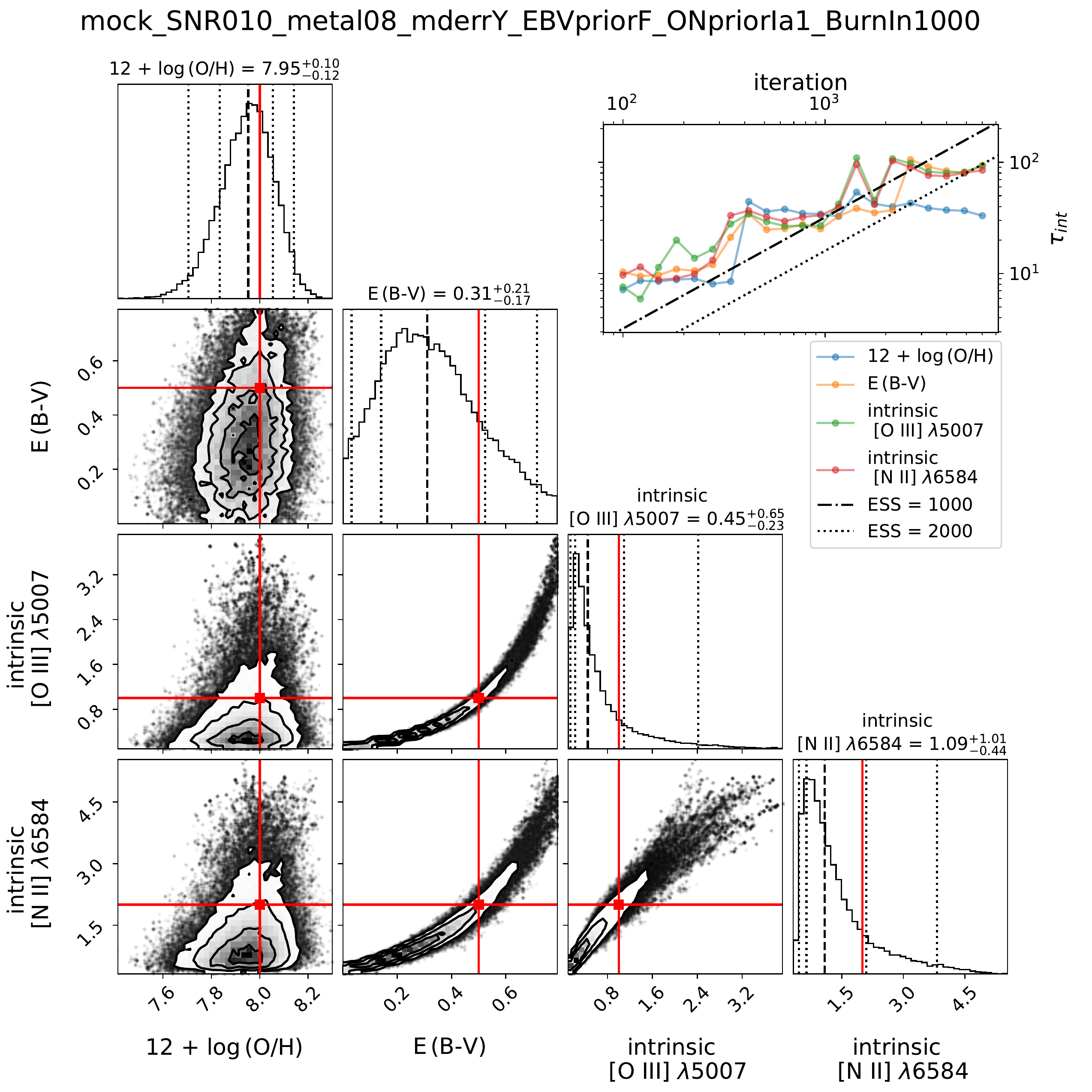}
    }
    \caption{Posterior distribution and evolution of the integrated autocorrelation times (\tauint) based on a mock-data test. Ten panels in the lower-left corner show the corner plots of the Markov chain Monte Carlo (MCMC) sampling results, demonstrating the correlations among the model parameters and their marginal distributions. `Intrinsic [O III] $\lambda5007$' and `intrinsic [N II] $\lambda6584$' indicate the corresponding emission-line fluxes in units of $10^{-16}$ \flux. Solid red lines represent input values used to generate the mock data. \del{Five vertical dashed lines}\add{One vertical dashed and four vertical dotted lines} indicate the median, 1-$\sigma$ (68 \%) and 2-$\sigma$ (95 \%) credible intervals\add{, respectively}. The panel in the upper-right corner shows the evolution of \tauint. For convergence diagnosis, I plot two straight lines (dot-dashed and dotted), corresponding to effective sample sizes (ESSs) of 1000 and 2000, respectively. The title at the top of the figure indicates the following---SNR\#\#\#: signal-to-noise ratio of the mock data, metal08: metallicity \metal{} = 8.0, mderrY: model uncertainty included, EBVpriorF: flat prior for \EBV{}, ONpriorIa1: inverse gamma prior for \OIIIint{} and intrinsic \NII{} line fluxes and BurnIn\#\#\#\#: burn-in iterations excluded before plotting. The `a1' suffix in the tag `ONpriorIa1' indicates that the priors use the shape parameter $a=1$. The complete figure set [nine images, i.e. combinations of input \EBV{} = $(0.1, 0.3 \textrm{ and } 0.5)$ and priors for \OIIIint{} and intrinsic \NII{} line fluxes = (flat, inverse gamma with $a=1$ and inverse gamma with $a=2$) = (ONpriorF, ONpriorIa1 and ONpriorIa2)] is available in the online journal.}
    \label{fig-mock-M}
\end{figure*}

The Bayesian parameter estimation results are shown in Figs.~\ref{fig-mock-M} and \ref{fig-cmp} and Table \ref{tbl-results}.
The lower-left part of Fig.~\ref{fig-mock-M} shows the corner plot, demonstrating the marginalised PDFs of the model parameters and the mutual relations for each parameter pair.
The upper-right part of Fig.~\ref{fig-mock-M} shows the evolution of \tauint{} along with two ESS guidelines, where the degree of the sampling convergence\footnote{It is crucial to monitor the convergence of MCMC sampling to the target distribution (here, the posterior) since it is based on a random process. I demonstrated the pitfall of the sampling without convergence monitoring in \cite{Shinn_2020_MNRAS_499_1073}.} can be observed.
In Fig.~\ref{fig-mock-M}, I only display the plots for the mock-data test with the input \EBV{} = 0.5 and inverse gamma priors with  $a=1$ for \OIIIint{} and intrinsic \NII{} line fluxes.
The results for all nine cases are available online; the nine cases are combinations of three \EBV{}s (0.1, 0.3 and 0.5) and three prior types for \OIIIint{} and intrinsic \NII{} line fluxes (`flat', `inverse gamma with $a=1$' and `inverse gamma with $a=2$').
The shapes of marginalised PDFs are similar for the inverse gamma prior cases, while those of the flat prior case are different.
This distinction is possibly attributed to the difference in the prior shapes.
The plots for the mutual relations between the model parameters are similar among the nine cases, although the high-density regions vary among cases.
The \tauint{} evolution plots show that the \tauint{}s of all parameters exceed the ESS of 2000 as I imposed for the sampling convergence.

Fig.~\ref{fig-cmp} shows each parameter's marginalised PDF and the corresponding input value for three prior cases considering the input \EBV{} = 0.5.
The other six cases for the input \EBV{} = 0.1 and 0.3 are available online.
Note that I adopted the metallicity calibration reported by \cite{Maiolino_2008_A&A_488_463} in this section.
The only case in which all input values were recovered within the 1-$\sigma$ credible interval is when adopting the inverse gamma prior with $a=1$ for \OIIIint{} and intrinsic \NII{} line fluxes.
However, the positions where the input values fall over the PDFs vary as the input \EBV{} varies even for this best case.
The input values of \EBV{}, intrinsic \OIII{} line flux and intrinsic \NII{} line flux fall near the upper 1-$\sigma$ credible limits of the corresponding PDFs when the input \EBV{} = 0.5; the input values move towards the lower 1-$\sigma$ credible limits when \EBV{} = 0.3 and further down when \EBV{} = 0.1 (Fig.~\ref{fig-cmp} and its online figures).
However, the change in \metal{} is relatively minor compared to the other three parameters.
The flat prior and inverse gamma prior with $a=2$ fail to recover all input values within the 1-$\sigma$ credible interval because they do not adequately handle the uncertainty term embedded in the likelihood [eq.~(\ref{eq-modL})].
Table \ref{tbl-results} summarises how the input values are recovered for all nine cases.
Again, all input values are recovered within the 1-$\sigma$ credible interval only when the inverse gamma prior with $a=1$ is adopted for \OIIIint{} and intrinsic \NII{} line fluxes.

\subsection{Mock-Data Test Based on a More Recent Metallicity Calibration} \label{ana-res-cal}
In section \ref{ana-res-mock}, I perform the mock-data tests based on the metallicity calibration reported by \cite{Maiolino_2008_A&A_488_463}.
However, as \cite{Indahl_2019_ApJ_883_114} and \cite{Shinn_2020_MNRAS_499_1073} mentioned, an improved metallicity calibration exists.
\cite{Curti_2017_MNRAS_465_1384} and \cite{Curti_2020_MNRAS_491_944} improved the calibration reported by \cite{Maiolino_2008_A&A_488_463} by uniformly applying the electron temperature method, which does not depend on modelling and is more direct, over the entire metallicity range.
To measure the electron temperatures at high metallicity, they stacked more than 110 000 galaxies from the \sdss{} Data Release 7 \citep[][]{Abazajian_2009_ApJS_182_543} in bins of log([O II]/H$\beta$) and log([O III]/H$\beta$).
The auroral emission lines used for the electron temperature measurement are usually challenging to detect from a single galaxy of high metallicity (e.g.~\metal{} $\goa$ 8.5); hence, they stacked the spectra of multiple galaxies.
I obtain almost the same results in this subsection when adopting a more recent metallicity calibration.

\begin{figure*}
    \center{
        \includegraphics[scale=0.55]{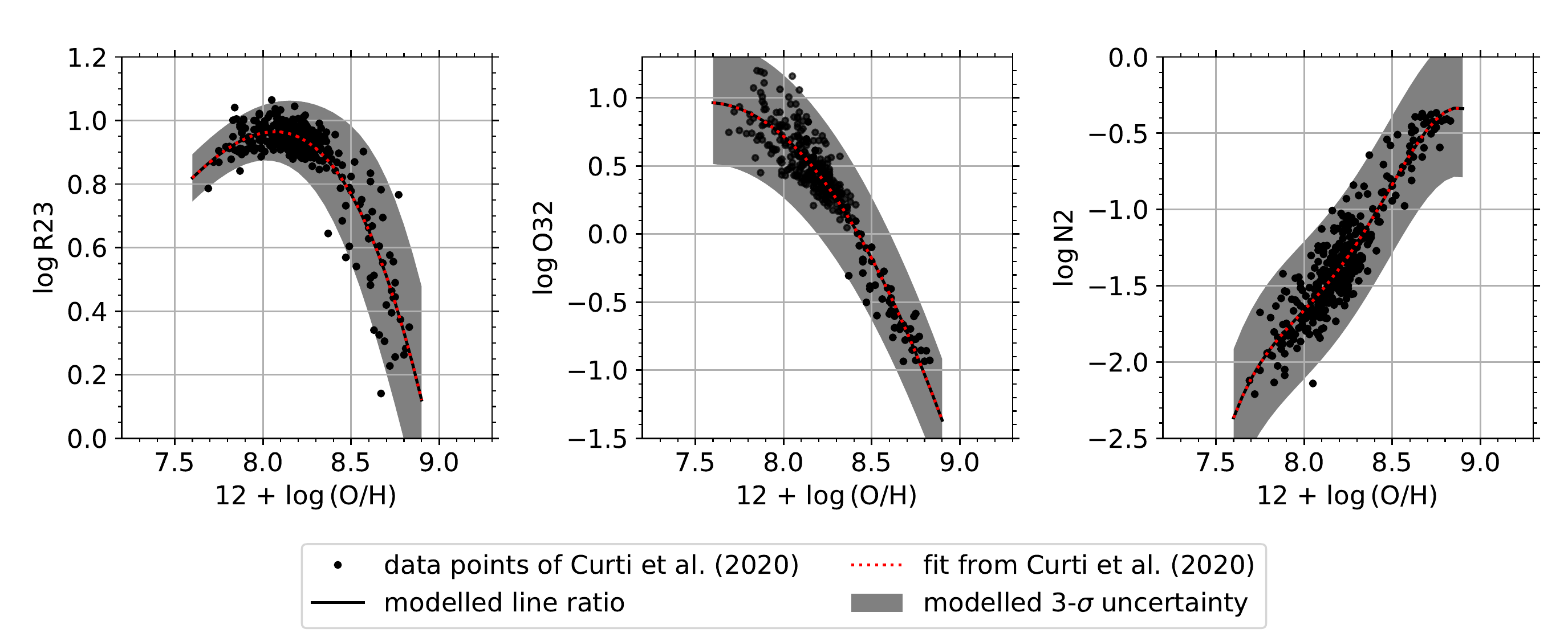}
    }
    \caption{Modelled line ratios and their uncertainties as a function of metallicity. Line ratios are defined in eqs. (\ref{R23})--(\ref{N2}). Black dots represent data points used by \protect\cite{Curti_2020_MNRAS_491_944} when deriving their metallicity calibration function (dotted red lines). Solid black line represents the line ratio calculated from the emission-line fluxes modelled in this work based on the metallicity calibration function reported by \protect\cite{Curti_2020_MNRAS_491_944}. Dotted red line and solid black line overlap entirely. Grey shading indicates the modelled 3-$\sigma$ uncertainty region, calculated from the emission-line fluxes' modelled uncertainties.}
    \label{fig-model_test}
\end{figure*}

I modelled the emission-line fluxes by replacing the metallicity calibration reported by \cite{Maiolino_2008_A&A_488_463} with \cite{Curti_2020_MNRAS_491_944}\footnote{The metallicity calibrations for the emission lines used in this study [eq.~(\ref{R23})-(\ref{N2})] are the same in both \cite{Curti_2017_MNRAS_465_1384} and \cite{Curti_2020_MNRAS_491_944}.}; except for the metallicity calibration, the rest of the modelling is the same as that described in section \ref{model-EBV}.
To assess the modelling of the emission-line fluxes, I prepare plots of metallicity as a function of the line ratios used in the modelling [eq.~(\ref{R23})-(\ref{N2})], and Fig.~\ref{fig-model_test} shows the results.
The line ratios of the modelled emission-line fluxes well follow the calibration reported by \cite{Curti_2020_MNRAS_491_944}.
I adjusted the line-ratio uncertainty to cover most of the data points that \cite{Curti_2020_MNRAS_491_944} employed at the 3-$\sigma$ level.
The uncertainties I adopted are as follows: $\Delta(\text{log\,O32})=0.15$, $\Delta(\text{log\,R23})=0.02$ and $\Delta(\text{log\,N2})=0.15$.
In addition, I set the arbitrary correlation between ($\text{[O II]}\,\lambda3727 + \text{[O III]}\,\lambda4959 + \text{[O III]}\,\lambda5007$) and R23 to $-1$ as done in \cite{Shinn_2020_MNRAS_499_1073}.
As Fig.~\ref{fig-model_test} shows, the line-ratio uncertainties adequately cover the spread of \citeauthor{Curti_2020_MNRAS_491_944}'s data points.

\begin{figure*}
    \center{
        \includegraphics[scale=0.55]{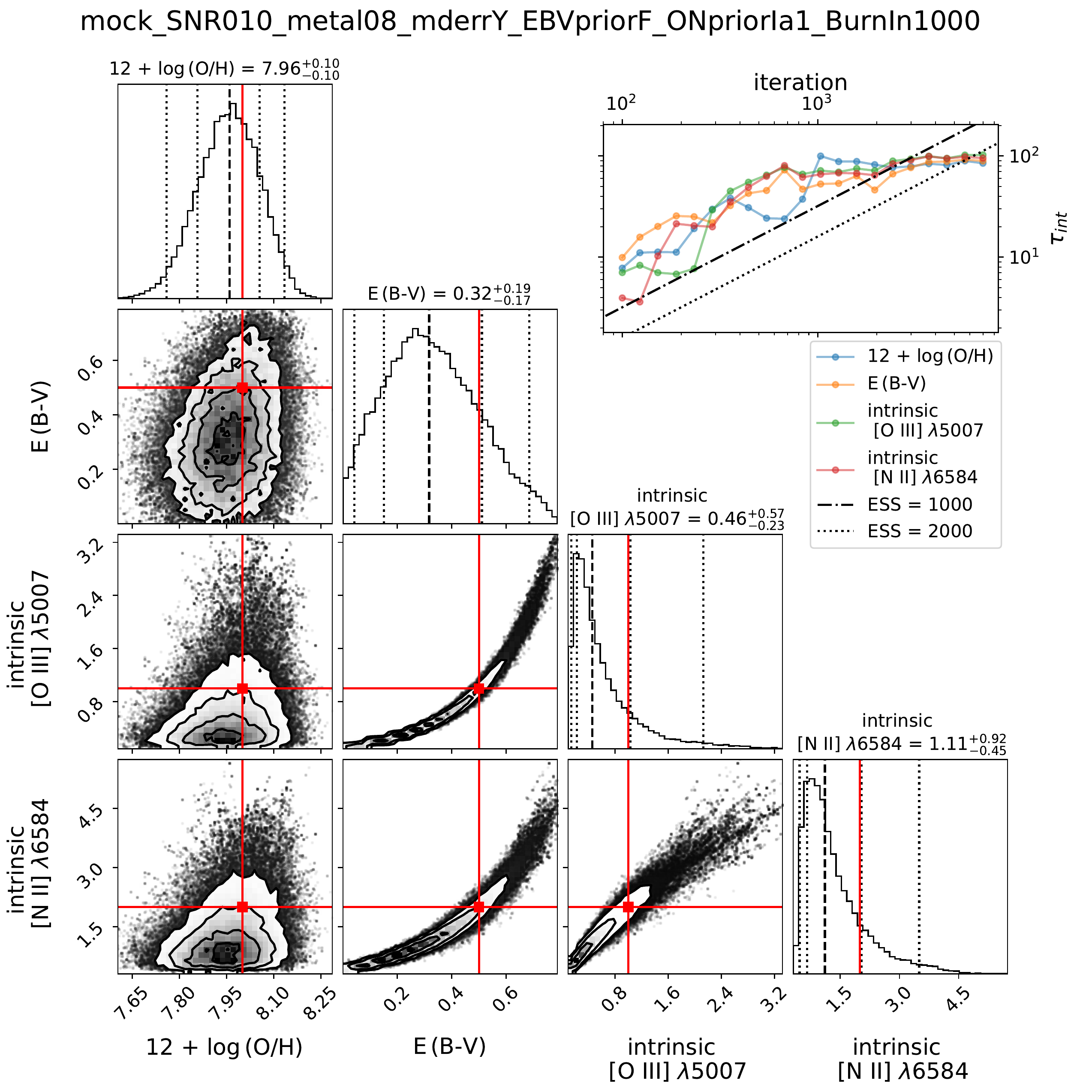}
    }
    \caption{Posterior distribution and evolution of the integrated autocorrelation times (\tauint) based on a mock-data test. The setting for this test is the same as that shown in Fig.~\ref{fig-mock-M}, except that the metallicity calibration function reported by \protect\cite{Curti_2020_MNRAS_491_944} is used to model the emission-line fluxes instead of that reported by \protect\cite{Maiolino_2008_A&A_488_463} (Fig.~\ref{fig-mock-M}). The figure description is otherwise the same as that of Fig.~\ref{fig-mock-M}. The complete figure set [nine images, i.e. combinations of \EBV{} = $(0.1, 0.3 \textrm{ and } 0.5)$ and priors for \OIIIint{} and intrinsic \NII{} line fluxes = (flat, inverse gamma with $a=1$ and inverse gamma with $a=2$) = (ONpriorF, ONpriorIa1 and ONpriorIa2)] is available in the online journal.}
    \label{fig-mock-C}
\end{figure*}

\begin{figure*}
    \center{
        \includegraphics[scale=0.6]{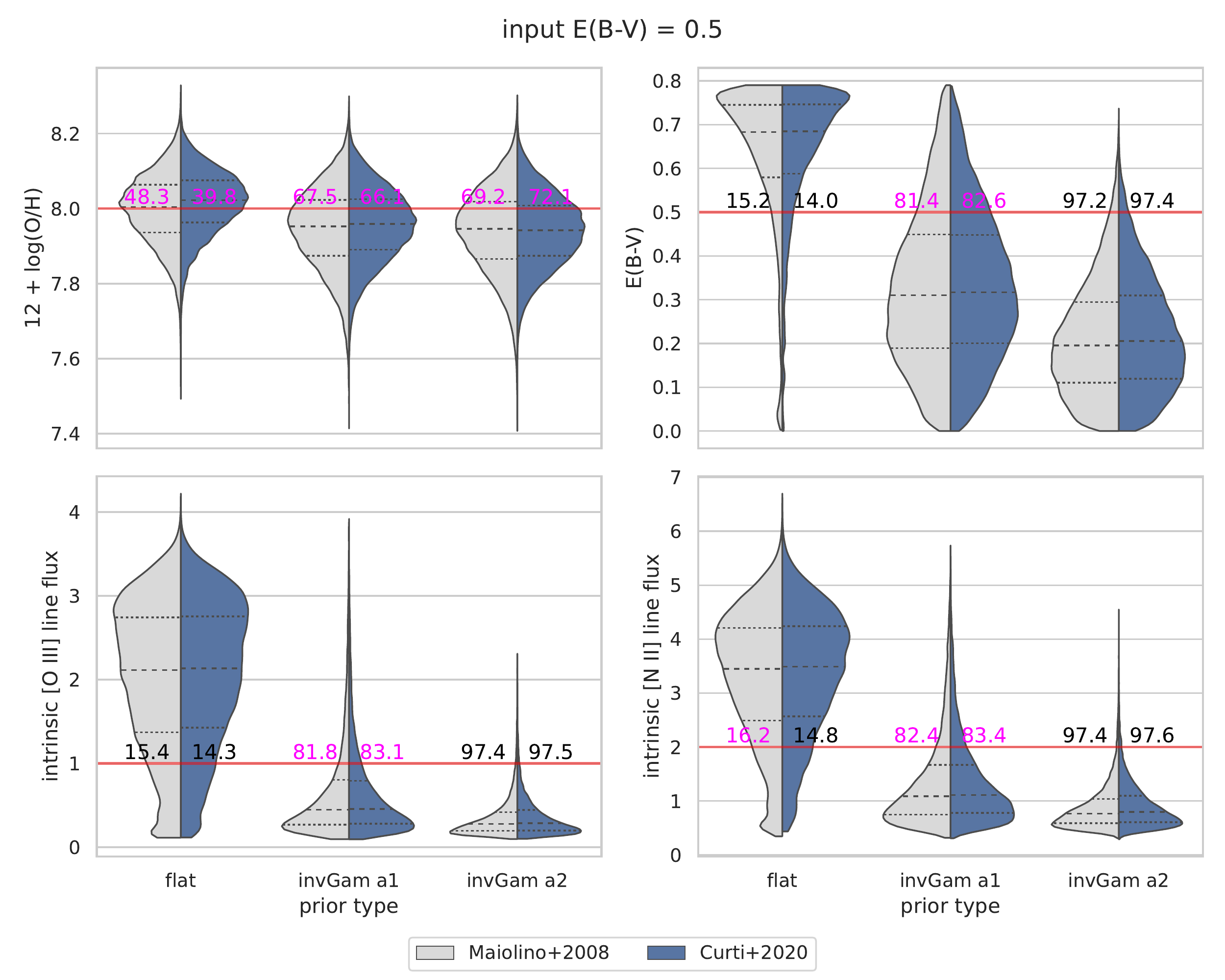}
    }
    \caption{Comparison between the parameter posterior distributions for the mock-data test with \EBV{} = 0.5. The abscissas represent the prior types for \OIIIint{} and intrinsic \NII{} line fluxes, and the ordinates represent the model parameters. The `flat', `invGam a1' and `invGam a2' indicate the flat, inverse gamma with $a=1$ and inverse gamma with $a=2$ priors, respectively. The shadings are different for different metallicity calibration functions used to model the emission-line fluxes. Horizontal black dashed and dotted lines indicate \add{the }median and quartile of the distribution, respectively. Horizontal red lines indicate input values for the mock data. The cumulative percentage at the input values, i.e. $P(y\le y_{\textrm{in}})$, is denoted on the horizontal red lines to show how well the posterior recover the input values. The number is indicated in magenta when the input value falls within the 1-$\sigma$ credible interval, i.e. $15.865<P(y\le y_{\textrm{in}})<84.135$. The complete figure set [three images, i.e. cases of \EBV{} = $(0.1, 0.3 \textrm{ and } 0.5)$] is available in the online journal.}
    \label{fig-cmp}
\end{figure*}

Using this model, I repeated the mock-data tests presented in section \ref{ana-res-mock}.
I slightly narrowed the range of the metallicity prior to make it consistent with the calibration range reported by \cite{Curti_2020_MNRAS_491_944} (Table \ref{tbl-priors}).
The results are presented in Figs.~\ref{fig-mock-C} and \ref{fig-cmp} and Table \ref{tbl-results}.
Fig.~\ref{fig-mock-C} shows the result for the same mock data used in Fig.~\ref{fig-mock-M}.
The results for all nine cases are available online, as in section \ref{ana-res-mock}.
The characteristics of the posterior distributions are almost the same as observed in the case of the results obtained using the metallicity calibration reported by \cite{Maiolino_2008_A&A_488_463}.
Fig.~\ref{fig-cmp} manifests such similarities.
Fig.~\ref{fig-cmp} shows each parameter's marginalised PDF and the corresponding input value for three prior cases for the input \EBV{} = 0.5.
The other six cases for the input \EBV{} = 0.1 and 0.3 are available online.
All cases show slight differences between the results obtained using the metallicity calibrations reported by \cite{Maiolino_2008_A&A_488_463} and \cite{Curti_2020_MNRAS_491_944}.
As in the case of \cite{Maiolino_2008_A&A_488_463}, the only scenario in which all input values are recovered within the 1-$\sigma$ credible interval is when adopting the inverse gamma prior with $a=1$ for \OIIIint{} and intrinsic \NII{} line fluxes.
Table \ref{tbl-results} summarises the recovery of the input values for all nine cases.
Again, all input values are recovered within the 1-$\sigma$ credible interval only when the inverse gamma prior with $a=1$ is adopted for \OIIIint{} and intrinsic \NII{} line fluxes.
Furthermore, slight differences were between the parameter estimations of the cases of \cite{Maiolino_2008_A&A_488_463} and \cite{Curti_2020_MNRAS_491_944}.

\section{Discussion}
Table \ref{tbl-results} summarises the results of the mock-data tests performed in this study, and Fig.~\ref{fig-cmp} shows the results\del{; all three figures for the cases of input \EBV{} = 0.1, 0.3 and 0.5 are available in the online journal}.
\add{The posteriors based on different priors and metallicity calibrations are plotted in parallel for all four model parameters in Fig.~\ref{fig-cmp} to show how well the input values (red horizontal lines) are recovered. The abscissa and ordinate denote three different priors and the four model parameters, respectively. Two different shades indicate two different metallicity calibrations. All the test results for the cases of input \EBV{} = 0.1, 0.3 and 0.5 are available in the online journal.}

Fig.~\ref{fig-cmp} shows that all four model parameters are recovered within the 1-$\sigma$ credible intervals only when the inverse gamma prior with $a=1$ is adopted for \OIIIint{} and intrinsic \NII{} line fluxes.
This result is interesting because I adopted the inverse gamma priors based on a logical motive: the intrinsic emission-line fluxes must exceed the observed (reddened) ones.
The results are even more contrasting when this case is compared to the flat prior case, which induces the \EBV{} overestimation and leads to the SFR overestimation (section \ref{model-EBV}).
The input \EBV{} value falls outside the 1-$\sigma$ credible interval for the input \EBV{} = 0.5 case and falls further outside the 2-$\sigma$ credible interval for the input \EBV{} = 0.1 case (the flat prior case of Fig.~\ref{fig-cmp} and online figures).
The flat prior case also overestimates \OIIIint{} and intrinsic \NII{} line fluxes in addition to \EBV{} because the model uncertainty considerably displaces the likelihood from where it should be [eq.~(\ref{eq-modL}) as well as Figs.~\ref{fig-mock-M} and \ref{fig-mock-C} and their online supplementary figures].
The \EBV{} overestimation is inevitable even when the normal (Gaussian) prior is adopted for \EBV{}, as reported previously \citep{Shinn_2020_MNRAS_499_1073}; in that study, flat priors were adopted for \OIIIint{} and intrinsic \NII{} line fluxes and the two parameters suffered from the 2-$\sigma$ overestimation (see Figs.~6-8 of the literature \citealt{Shinn_2020_MNRAS_499_1073}).
Therefore, it is remarkable that the \EBV{} overestimation problem in \cite{Shinn_2020_MNRAS_499_1073} can be solved when using the inverse gamma prior with $a=1$ for \OIIIint{} and intrinsic \NII{} line fluxes, and the priors are set to have their maxima at the corresponding observed emission-line fluxes (section \ref{ana-res-mock}).
The independent determination of \EBV{}, which was proposed for solving the \EBV{} overestimation problem \citep{Shinn_2020_MNRAS_499_1073}, is still the most reliable solution because the \EBV{} value is measured directly.
However, this independent \EBV{} determination requires an additional observation time.
Therefore, the proposed prior treatment is beneficial because improved parameter estimates can be obtained without additional observations.

I performed the mock-data test using the two metallicity calibrations reported by \cite{Maiolino_2008_A&A_488_463} and \cite{Curti_2020_MNRAS_491_944}.
Fig.~\ref{fig-cmp} and its online supplementary figures show that the two calibrations present slight differences in the recovery of the input values, probably because the two calibrations show similar shapes and spreading for the line ratios as a function of metallicity.
A similar shape can be observed in the comparison plots presented by \cite{Curti_2017_MNRAS_465_1384}.
When using the two calibrations, I adopted a flat prior for the metallicity and limited the range to be the valid range of the corresponding calibration (Table \ref{tbl-priors}).
The overlapping range for the two calibrations is $7.6-8.9$.
I set \metal{} = 8.0 for the mock-data test (Table \ref{tbl-results}).
To assess whether the input values are successfully recovered under different metallicity conditions, I repeated the mock-data test for \metal{} = 8.5.
I found that the posterior's properties are almost the same, and the recovery is marginally successful; the input values of \EBV{}, intrinsic \OIII{} line flux and intrinsic \NII{} line flux fall slightly outside the upper 1-$\sigma$ credible limit.

\begin{figure*}
    \center{
        \includegraphics[scale=0.6]{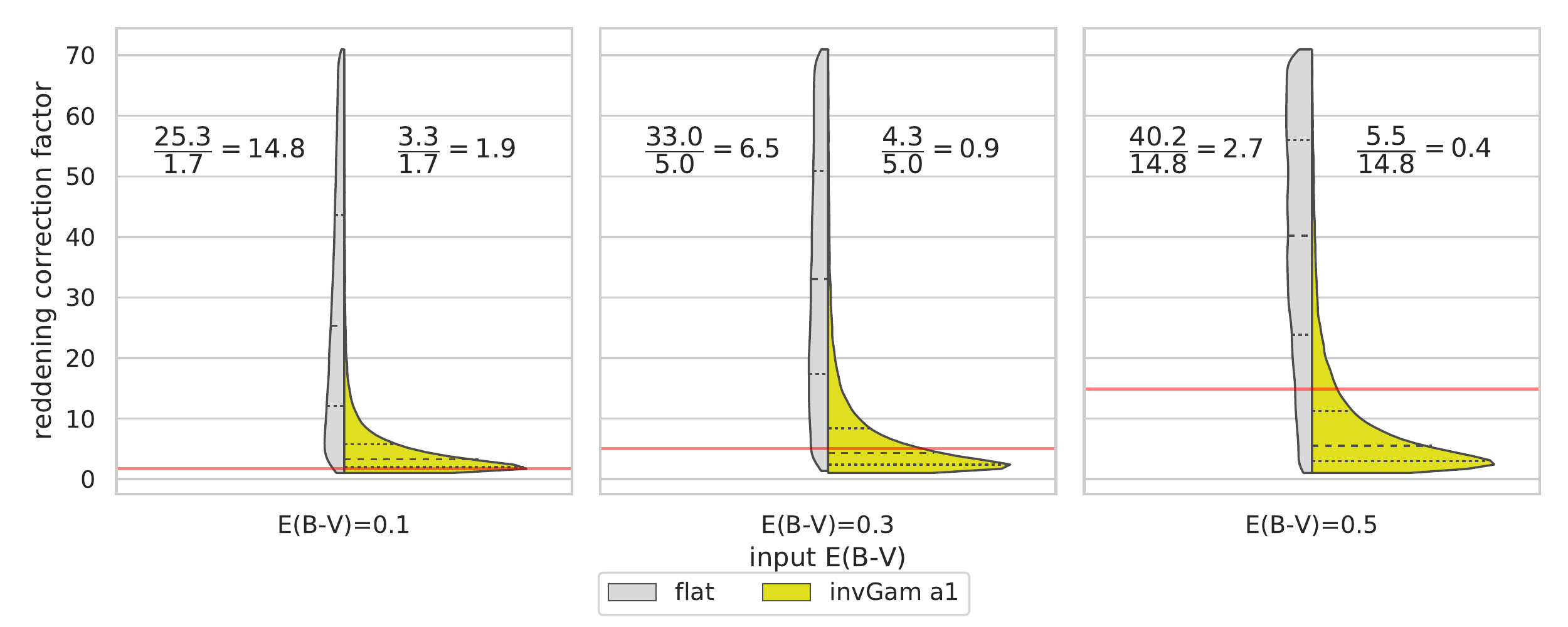}
    }
    \caption{\add{Comparison between the reddening correction factors for the [O II] $\lambda3727$ line flux. Each panel compares the correction factors derived from two \EBV{} posteriors obtained from the mock data tests by using the Calzetti attenuation curve \protect\citep[][]{Calzetti_2000_ApJ_533_682}. The two \EBV{} posteriors are from the mock data test results with two different priors for \OIIIint{} and \NIIint{} line fluxes: flat (`flat') and inverse gamma with $a=1$ (`invGam a1'). This prior difference is indicated with different shadings (grey and yellow). Three panels correspond to the mock data test results with three different input \EBV{} values. The presented mock data test results are based on the metallicity calibration function of \protect\cite{Curti_2020_MNRAS_491_944}. Horizontal black dashed and dotted lines indicate the median and quartile of the distribution, respectively. Horizontal red lines indicate the correction factors corresponding to the input \EBV{} values. The printed number shows the ratio of the correction factor at the median to the one at the input \EBV{} value, i.e. $f_{\textrm{med}}/f_{\textrm{input}}$.}}
    \label{fig-cmp-cf}
\end{figure*}

\add{To show how much better SFR estimates we can have when the inverse gamma prior with $a=1$ is adopted for \OIIIint{} and \NIIint{} line fluxes instead of the flat prior, I compared the reddening correction factor for the [O II] $\lambda3727$ line flux, which is used for the SFR estimation.
\cite{Indahl_2019_ApJ_883_114} estimated the SFRs for the HPS target galaxies using two relations derived by \cite{Kewley_2004_AJ_127_2002}, both of which show a linear dependence of the SFR on the luminosity of [O II] $\lambda3727$ line ($\textrm{SFR}\propto L_{\textrm{[O II]}}$).
Therefore, the SFR estimates also linearly depend on the reddening correction factor for the [O II] $\lambda3727$ line flux, which subsequently depends on \EBV{}.
Using the \EBV{} posteriors obtained from the mock data tests with two different priors for \OIIIint{} and \NIIint{} line fluxes (flat and inverse gamma with $a=1$), I compared the correction factors.
I chose the mock data test results obtained with the metallicity calibration function of \cite{Curti_2020_MNRAS_491_944}.
Fig.~\ref{fig-cmp-cf} shows the results for three different input \EBV{} values.
For all three input \EBV{} cases, the inverse gamma with $a=1$ prior case presents a better recovery of the correction factor at the input \EBV{} value.
The superiority of the inverse gamma with $a=1$ prior case is more evident when comparing the ratio of the correction factor at the median to the one at the input \EBV{} value, i.e.~$f_{\textrm{med}}/f_{\textrm{input}}$.
When the input \EBV{} = 0.5, $f_{\textrm{med}}/f_{\textrm{input}}$s deviate from one by similar factors (2.7 and $1/0.4=2.5$).
However, when the input \EBV{} = 0.1, $f_{\textrm{med}}/f_{\textrm{input}}$ of the inverse gamma with $a=1$ prior case (1.9) is about eight times smaller than the flat prior case (14.8).
This difference means that the inverse gamma with $a=1$ prior case can diminish the SFR overestimation of the flat prior case eightfold.
}

The mock-data tests presented in this study show that the recovery of the input values can be substantially improved if the inverse gamma distribution with $a=1$ replaces the flat distribution for the priors of \OIIIint{} and intrinsic \NII{} line fluxes (Fig.~\ref{fig-cmp} and its online supplementary figures).
The inverse gamma distribution reflects the logical constraint that an intrinsic emission-line flux must exceed the observed (reddened) emission-line flux.
This outcome implies that the recovery of the input values can be improved if the given constraints are appropriately shaped into the priors when performing the Bayesian parameter estimation.
This type of treatment is impossible in the classical frequentist approach for the parameter estimation because it only uses the likelihood; there is no means to correct the parameter estimation once the likelihood is substantially displaced owing to the uncertainty term.
Therefore, I herein demonstrate how the Bayesian parameter estimation becomes an alternative when the classical parameter estimation becomes a challenging problem.
Moreover, the findings of this study indicate that the additional observation time can be saved, which may be used for solving the problem caused by the displaced likelihood, similar to the extra observation time for the independent \EBV{} determination proposed in \cite{Shinn_2020_MNRAS_499_1073}.

\section{Conclusion}
My reanalysis of the local star-forming galaxies observed in the HPS \citep[][]{Shinn_2020_MNRAS_499_1073} showed that the discovery of a new galaxy population claimed in the original study of \cite{Indahl_2019_ApJ_883_114} is probably based on the overestimated SFRs, which stemmed from the \EBV{} overestimation during their Bayesian parameter estimation.
\cite{Indahl_2019_ApJ_883_114} and \cite{Shinn_2020_MNRAS_499_1073} modelled the observed emission-line fluxes using the strong-line method with four parameters: metallicity \metal{}, nebula emission-line colour excess \EBV{}, intrinsic \OIII{} line flux and intrinsic \NII{} line flux.
The \EBV{} overestimation was attributed to the uncertainty of the metallicity calibration, which enters the likelihood and yields inaccurate parameter estimates.

In this study, I determine whether new scientifically motivated proper priors can alleviate the \EBV{} overestimation problem.
I found that the problem can be eased if suitable priors are adopted.
I adopted inverse gamma distributions for \OIIIint{} and intrinsic \NII{} line fluxes and set their mode equal to the observed emission-line fluxes.
This configuration reflects the logical constraint that an intrinsic emission-line flux must exceed the observed (reddened) one.
Based on the mock-data tests, the input values can be recovered within and around the 1-$\sigma$ credible interval when using the inverse gamma distributions with $a=1$ for the priors of \OIIIint{} and intrinsic \NII{} line fluxes.
Similar results are obtained when using either of the two metallicity calibrations [\cite{Maiolino_2008_A&A_488_463} and \cite{Curti_2020_MNRAS_491_944}], either of the three colour excess input values [\EBV{} = 0.1, 0.3 and 0.5] and either of the two metallicity input values [\metal{} = 8.0 and 8.5].
\add{I also show that the SFR overestimation can be decreased eightfold when the inverse gamma with $a=1$ priors are adopted instead of the flat priors.}
This study thus suggests that more accurate estimates for the model parameters are obtainable if the given constraints are appropriately included in the priors during the Bayesian parameter estimation, particularly when the likelihood does not correctly constraint the model parameters, i.e. the case where the classical frequentist parameter estimation approach becomes challenging.
This treatment is beneficial because no further observations are needed to correct the inaccurate classical parameter estimates.

\section*{Acknowledgements}
The author appreciates the comments from the anonymous referee, which significantly improve the draft.
The author is also grateful to Mirko Curti for kindly providing the metallicity calibration data and verifying the calibration plots.
This research was supported by the Korea Astronomy and Space Science Institute under the R\&D program(Project No. 2022-1-868-04), supervised by the Ministry of Science and ICT.

\section*{Data Availability}


All MCMC sampling results can be downloaded from \url{https://data.kasi.re.kr/vo/Stat_Reanal/} with a plotting script.



\bibliographystyle{mnras}
\bibliography{stat_reanal_2020a} 

\begin{thebibliography}{}
\makeatletter
\relax
\def\mn@urlcharsother{\let\do\@makeother \do\$\do\&\do\#\do\^\do\_\do\%\do\~}
\def\mn@doi{\begingroup\mn@urlcharsother \@ifnextchar [ {\mn@doi@}
  {\mn@doi@[]}}
\def\mn@doi@[#1]#2{\def\@tempa{#1}\ifx\@tempa\@empty \href
  {http://dx.doi.org/#2} {doi:#2}\else \href {http://dx.doi.org/#2} {#1}\fi
  \endgroup}
\def\mn@eprint#1#2{\mn@eprint@#1:#2::\@nil}
\def\mn@eprint@arXiv#1{\href {http://arxiv.org/abs/#1} {{\tt arXiv:#1}}}
\def\mn@eprint@dblp#1{\href {http://dblp.uni-trier.de/rec/bibtex/#1.xml}
  {dblp:#1}}
\def\mn@eprint@#1:#2:#3:#4\@nil{\def\@tempa {#1}\def\@tempb {#2}\def\@tempc
  {#3}\ifx \@tempc \@empty \let \@tempc \@tempb \let \@tempb \@tempa \fi \ifx
  \@tempb \@empty \def\@tempb {arXiv}\fi \@ifundefined
  {mn@eprint@\@tempb}{\@tempb:\@tempc}{\expandafter \expandafter \csname
  mn@eprint@\@tempb\endcsname \expandafter{\@tempc}}}

\bibitem[\protect\citeauthoryear{Abazajian et~al.,}{Abazajian
  et~al.}{2009}]{Abazajian_2009_ApJS_182_543}
Abazajian K.~N.,  et~al., 2009, \mn@doi [ApJS] {10.1088/0067-0049/182/2/543},
  182, 543

\bibitem[\protect\citeauthoryear{{Adams} et~al.,}{{Adams}
  et~al.}{2011}]{Adams_2011_ApJS_192_5}
{Adams} J.~J.,  et~al., 2011, \mn@doi [ApJS] {10.1088/0067-0049/192/1/5}, 192,
  5

\bibitem[\protect\citeauthoryear{Birrer et~al.,}{Birrer
  et~al.}{2019}]{Birrer_2019_MNRAS_484_4726}
Birrer S.,  et~al., 2019, \mn@doi [MNRAS] {10.1093/mnras/stz200}, 484, 4726

\bibitem[\protect\citeauthoryear{Brooks, Gelman, Jones  \& Meng}{Brooks
  et~al.}{2011}]{Brooks_2011_book}
Brooks S.,  Gelman A.,  Jones G.,   Meng X.-L.,  eds, 2011, Handbook of Markov
  Chain Monte Carlo (Chapman \& Hall/CRC Handbooks of Modern Statistical
  Methods).
Chapman and Hall/CRC (Boca Raton, FL 33487, USA)

\bibitem[\protect\citeauthoryear{Brown, Latham, Everett  \& Esquerdo}{Brown
  et~al.}{2011}]{Brown_2011_AJ_142_112}
Brown T.~M.,  Latham D.~W.,  Everett M.~E.,   Esquerdo G.~A.,  2011, \mn@doi
  [AJ] {10.1088/0004-6256/142/4/112}, 142, 112

\bibitem[\protect\citeauthoryear{Buchner et~al.,}{Buchner
  et~al.}{2014}]{Buchner_2014_A&A_564_A125}
Buchner J.,  et~al., 2014, \mn@doi [A\&A] {10.1051/0004-6361/201322971}, 564,
  A125

\bibitem[\protect\citeauthoryear{{Calzetti}, {Armus}, {Bohlin}, {Kinney},
  {Koornneef}  \& {Storchi-Bergmann}}{{Calzetti}
  et~al.}{2000}]{Calzetti_2000_ApJ_533_682}
{Calzetti} D.,  {Armus} L.,  {Bohlin} R.~C.,  {Kinney} A.~L.,  {Koornneef} J.,
   {Storchi-Bergmann} T.,  2000, \mn@doi [ApJ] {10.1086/308692}, 533, 682

\bibitem[\protect\citeauthoryear{Cameron, Angus  \& Burgess}{Cameron
  et~al.}{2020}]{Cameron_2020_NatAs_4_132}
Cameron E.,  Angus G.~W.,   Burgess J.~M.,  2020, \mn@doi [NatAs]
  {10.1038/s41550-019-0998-2}, 4, 132

\bibitem[\protect\citeauthoryear{{Chonis} et~al.,}{{Chonis}
  et~al.}{2016}]{Chonis_2016_inbook}
{Chonis} T.~S.,  et~al., 2016, {LRS2: design, assembly, testing, and
  commissioning of the second-generation low-resolution spectrograph for the
  Hobby-Eberly Telescope}.
p. 99084C, \mn@doi{10.1117/12.2232209}, \url
  {https://ui.adsabs.harvard.edu/abs/2016SPIE.9908E..4CC}

\bibitem[\protect\citeauthoryear{{Curti}, {Cresci}, {Mannucci}, {Marconi},
  {Maiolino}  \& {Esposito}}{{Curti} et~al.}{2017}]{Curti_2017_MNRAS_465_1384}
{Curti} M.,  {Cresci} G.,  {Mannucci} F.,  {Marconi} A.,  {Maiolino} R.,
  {Esposito} S.,  2017, \mn@doi [MNRAS] {10.1093/mnras/stw2766}, 465, 1384

\bibitem[\protect\citeauthoryear{Curti, Mannucci, Cresci  \& Maiolino}{Curti
  et~al.}{2020}]{Curti_2020_MNRAS_491_944}
Curti M.,  Mannucci F.,  Cresci G.,   Maiolino R.,  2020, \mn@doi [MNRAS]
  {10.1093/mnras/stz2910}, 491, 944

\bibitem[\protect\citeauthoryear{Díaz, Almenara, Santerne, Moutou, Lethuillier
   \& Deleuil}{Díaz et~al.}{2014}]{Diaz_2014_MNRAS_441_983}
Díaz R.~F.,  Almenara J.~M.,  Santerne A.,  Moutou C.,  Lethuillier A.,
  Deleuil M.,  2014, \mn@doi [MNRAS] {10.1093/mnras/stu601}, 441, 983

\bibitem[\protect\citeauthoryear{D’Antona, Caloi  \& Tailo}{D’Antona
  et~al.}{2018}]{DAntona_2018_NatAs__}
D’Antona F.,  Caloi V.,   Tailo M.,  2018, NatAs

\bibitem[\protect\citeauthoryear{Farr, Sravan, Cantrell, Kreidberg, Bailyn,
  Mandel  \& Kalogera}{Farr et~al.}{2011}]{Farr_2011_ApJ_741_103}
Farr W.~M.,  Sravan N.,  Cantrell A.,  Kreidberg L.,  Bailyn C.~D.,  Mandel I.,
    Kalogera V.,  2011, \mn@doi [ApJ] {10.1088/0004-637X/741/2/103}, 741, 103

\bibitem[\protect\citeauthoryear{{Foreman-Mackey}, {Hogg}, {Lang}  \&
  {Goodman}}{{Foreman-Mackey} et~al.}{2013}]{Foreman-Mackey_2013_PASP_125_306}
{Foreman-Mackey} D.,  {Hogg} D.~W.,  {Lang} D.,   {Goodman} J.,  2013, \mn@doi
  [PASP] {10.1086/670067}, 125, 306

\bibitem[\protect\citeauthoryear{{Foreman-Mackey} et~al.,}{{Foreman-Mackey}
  et~al.}{2019}]{Foreman-Mackey_2019_JOSS_4_1864}
{Foreman-Mackey} D.,  et~al., 2019, \mn@doi [JOSS] {10.21105/joss.01864}, 4,
  1864

\bibitem[\protect\citeauthoryear{Goodman \& Weare}{Goodman \&
  Weare}{2010}]{Goodman_2010_CAMCoS_5_65}
Goodman J.,  Weare J.,  2010, CAMCoS, 5, 65

\bibitem[\protect\citeauthoryear{{Grasshorn Gebhardt}, {Zeimann}, {Ciardullo},
  {Gronwall}, {Hagen}, {Bridge}, {Schneider}  \& {Trump}}{{Grasshorn Gebhardt}
  et~al.}{2016}]{GrasshornGebhardt_2016_ApJ_817_10}
{Grasshorn Gebhardt} H.~S.,  {Zeimann} G.~R.,  {Ciardullo} R.,  {Gronwall} C.,
  {Hagen} A.,  {Bridge} J.~S.,  {Schneider} D.~P.,   {Trump} J.~R.,  2016,
  \mn@doi [ApJ] {10.3847/0004-637X/817/1/10}, \href
  {https://ui.adsabs.harvard.edu/abs/2016ApJ...817...10G} {817, 10}

\bibitem[\protect\citeauthoryear{Gregory}{Gregory}{2005}]{Gregory_2005_ApJ_631_1198}
Gregory P.~C.,  2005, \mn@doi [ApJ] {10.1086/432594}, 631, 1198

\bibitem[\protect\citeauthoryear{{Hill} et~al.,}{{Hill}
  et~al.}{2008}]{Hill_2008_inproca}
{Hill} G.~J.,  et~al., 2008, in {Kodama} T.,  {Yamada} T.,   {Aoki} K.,  eds,
  Astronomical Society of the Pacific Conference Series Vol. 399, Panoramic
  Views of Galaxy Formation and Evolution. p.~115 (\mn@eprint {arXiv}
  {0806.0183})

\bibitem[\protect\citeauthoryear{Hill et~al.,}{Hill
  et~al.}{2021}]{Hill_2021_AJ_162_298}
Hill G.~J.,  et~al., 2021, \mn@doi [AJ] {10.3847/1538-3881/ac2c02}, 162, 298

\bibitem[\protect\citeauthoryear{Hobert \& Casella}{Hobert \&
  Casella}{1996}]{Hobert_1996_J.Am.Stat.Assoc._91_1461}
Hobert J.~P.,  Casella G.,  1996, \mn@doi [J. Am. Stat. Assoc.]
  {10.1080/01621459.1996.10476714}, 91, 1461

\bibitem[\protect\citeauthoryear{{Hogg} \& {Foreman-Mackey}}{{Hogg} \&
  {Foreman-Mackey}}{2018}]{Hogg_2018_ApJS_236_11}
{Hogg} D.~W.,  {Foreman-Mackey} D.,  2018, \mn@doi [ApJS]
  {10.3847/1538-4365/aab76e}, 236, 11

\bibitem[\protect\citeauthoryear{Indahl et~al.,}{Indahl
  et~al.}{2019}]{Indahl_2019_ApJ_883_114}
Indahl B.,  et~al., 2019, \mn@doi [ApJ] {10.3847/1538-4357/ab3df7}, 883, 114

\bibitem[\protect\citeauthoryear{{Kewley}, {Geller}  \& {Jansen}}{{Kewley}
  et~al.}{2004}]{Kewley_2004_AJ_127_2002}
{Kewley} L.~J.,  {Geller} M.~J.,   {Jansen} R.~A.,  2004, \mn@doi [AJ]
  {10.1086/382723}, \href
  {https://ui.adsabs.harvard.edu/abs/2004AJ....127.2002K} {127, 2002}

\bibitem[\protect\citeauthoryear{{Maiolino} \& {Mannucci}}{{Maiolino} \&
  {Mannucci}}{2019}]{Maiolino_2019_A&ARv_27_3}
{Maiolino} R.,  {Mannucci} F.,  2019, \mn@doi [A\&ARv]
  {10.1007/s00159-018-0112-2}, 27, 3

\bibitem[\protect\citeauthoryear{{Maiolino} et~al.,}{{Maiolino}
  et~al.}{2008}]{Maiolino_2008_A&A_488_463}
{Maiolino} R.,  et~al., 2008, \mn@doi [A\&A] {10.1051/0004-6361:200809678},
  \href {https://ui.adsabs.harvard.edu/abs/2008A&A...488..463M} {488, 463}

\bibitem[\protect\citeauthoryear{Martinez}{Martinez}{2015}]{Martinez_2015_MNRAS_451_2524}
Martinez G.~D.,  2015, \mn@doi [MNRAS] {10.1093/mnras/stv942}, 451, 2524

\bibitem[\protect\citeauthoryear{Miller, Kitching, Heymans, Heavens  \& van
  Waerbeke}{Miller et~al.}{2007}]{Miller_2007_MNRAS_382_315}
Miller L.,  Kitching T.~D.,  Heymans C.,  Heavens A.~F.,   van Waerbeke L.,
  2007, \mn@doi [MNRAS] {10.1111/j.1365-2966.2007.12363.x}, 382, 315

\bibitem[\protect\citeauthoryear{{Planck Collaboration} et~al.,}{{Planck
  Collaboration} et~al.}{2018}]{PlanckCollaboration_2018_arXiv_07_6211}
{Planck Collaboration} et~al., 2018, arXiv, 07, 6211

\bibitem[\protect\citeauthoryear{Raftery \& Lewis}{Raftery \&
  Lewis}{1992}]{Raftery_1992_inproc}
Raftery A.~E.,  Lewis S.,  1992, in Bayesian Statistics 4. Clarendon Press
  (Oxford, UK), pp 763--773

\bibitem[\protect\citeauthoryear{Ramsey et~al.,}{Ramsey
  et~al.}{1998}]{Ramsey_1998_inproc}
Ramsey L.~W.,  et~al., 1998, in Society of Photo-Optical Instrumentation
  Engineers (SPIE) Conference Series: Advanced Technology Optical/IR Telescopes
  VI. pp 34--42

\bibitem[\protect\citeauthoryear{Sharma}{Sharma}{2017}]{Sharma_2017_ARA&A_55_213}
Sharma S.,  2017, \mn@doi [ARA\&A] {10.1146/annurev-astro-082214-122339}, 55,
  213

\bibitem[\protect\citeauthoryear{{Shinn}}{{Shinn}}{2020}]{Shinn_2020_MNRAS_499_1073}
{Shinn} J.-H.,  2020, \mn@doi [MNRAS] {10.1093/mnras/staa2836}, \href
  {https://ui.adsabs.harvard.edu/abs/2020MNRAS.499.1073S} {499, 1073}

\bibitem[\protect\citeauthoryear{Sivia \& Skilling}{Sivia \&
  Skilling}{2006}]{Sivia_2006_book}
Sivia D.,  Skilling J.,  2006, Data Analysis: A Bayesian Tutorial.
Oxford University Press

\bibitem[\protect\citeauthoryear{Starck, Donoho, Fadili  \& Rassat}{Starck
  et~al.}{2013}]{Starck_2013_A&A_552_A133}
Starck J.-L.,  Donoho D.~L.,  Fadili M.~J.,   Rassat A.,  2013, \mn@doi [A\&A]
  {10.1051/0004-6361/201321257}, 552, A133

\bibitem[\protect\citeauthoryear{Steiner, Lattimer  \& Brown}{Steiner
  et~al.}{2010}]{Steiner_2010_ApJ_722_33}
Steiner A.~W.,  Lattimer J.~M.,   Brown E.~F.,  2010, \mn@doi [ApJ]
  {10.1088/0004-637X/722/1/33}, 722, 33

\bibitem[\protect\citeauthoryear{{Storey} \& {Zeippen}}{{Storey} \&
  {Zeippen}}{2000}]{Storey_2000_MNRAS_312_813}
{Storey} P.~J.,  {Zeippen} C.~J.,  2000, \mn@doi [MNRAS]
  {10.1046/j.1365-8711.2000.03184.x}, \href
  {https://ui.adsabs.harvard.edu/abs/2000MNRAS.312..813S} {312, 813}

\bibitem[\protect\citeauthoryear{Storn \& Price}{Storn \&
  Price}{1997}]{Storn_1997_J.GlobalOptim._11_341}
Storn R.,  Price K.,  1997, \mn@doi [J. Global Optim.]
  {10.1023/A:1008202821328}, 11, 341

\bibitem[\protect\citeauthoryear{Tak, Ghosh  \& Ellis}{Tak
  et~al.}{2018}]{Tak_2018_MNRAS_481_277}
Tak H.,  Ghosh S.~K.,   Ellis J.~A.,  2018, \mn@doi [MNRAS]
  {10.1093/mnras/sty2326}, 481, 277

\bibitem[\protect\citeauthoryear{Wasserman}{Wasserman}{2004}]{Wasserman_2004_book}
Wasserman L.,  2004, All of Statistics.
Springer-Verlag GmbH

\bibitem[\protect\citeauthoryear{Wolfgang \& Lopez}{Wolfgang \&
  Lopez}{2015}]{Wolfgang_2015_ApJ_806_183}
Wolfgang A.,  Lopez E.,  2015, \mn@doi [ApJ] {10.1088/0004-637X/806/2/183},
  806, 183

\bibitem[\protect\citeauthoryear{von Toussaint}{von
  Toussaint}{2011}]{Toussaint_2011_RvMP_83_943}
von Toussaint U.,  2011, \mn@doi [RvMP] {10.1103/RevModPhys.83.943}, 83, 943

\makeatother
\end{thebibliography}








\bsp	
\label{lastpage}
\end{document}